\documentclass{paper}
\usepackage[latin9]{inputenc}
\usepackage{color}
\usepackage{float}
\usepackage{amsmath}
\usepackage{graphicx}

\makeatletter

\floatstyle{ruled}
\newfloat{algorithm}{tbp}{loa}
\providecommand{\algorithmname}{Algorithm}
\floatname{algorithm}{\protect\algorithmname}

\makeatother

\begin{document}

\title{A Novel Protocol for Network-Controlled Metasurfaces}

\author{Angeliki Tsioliaridou\\
FORTH, Greece\\
atsiolia@ics.forth.gr\\
Christos Liaskos\\
FORTH, Greece\\
cliaskos@ics.forth.gr\\
Andreas Pitsillides\\
University of Cyprus, Cyprus\\
Andreas.Pitsillides@ucy.ac.cy\\
Sotiris Ioannidis\\
FORTH, Greece\\
sotiris@ics.forth.gr\thanks{This work was funded by the European Union via the Horizon 2020: Future
Emerging Topics call (FETOPEN), grant EU736876, project VISORSURF
(http://www.visorsurf.eu).}}
\maketitle
\begin{abstract}
A recently proposed class of materials, called software-defined metamaterials,
can change their electromagnetic behavior on demand, utilizing a nanonetwork
embedded in their structure. The present work focuses on 2D metamaterials,
known as metasurfaces, and their electromagnetically programmable
counterparts, the HyperSurfaces. The particular focus of the study
is to propose a nanonetworking protocol that can support the intended
macroscopic functionality of a HyperSurface, such as sensing and reacting
to impinging waves in a customizable manner. The novel protocol is
derived analytically, using the Lyapunov drift minimization approach,
taking into account nano-node energy, communication latency and complexity
concerns. The proposed scheme is evaluated via simulations, covering
both the macroscopic HyperSurface functionality and the microscopic,
nanonetwork behavior.
\end{abstract}
\keywords{Electromagnetic nano-networking, metasurfaces, protocol.}

\section{Introduction\label{sec:Introduction}}

Extending the reach of communications, control and networking to the
nano-meter level offers unprecedented applications in sectors such
as medicine and material manufacturing~\cite{Piro.2015,Akyildiz.2010c}.
Targeted drug delivery at cellular level and sensing for material
manufacturing flaws at nano-scale are among the envisioned tasks.
Recently, another application of nanonetworks has been proposed, in
the form of objects with programmable electromagnetic behavior~\cite{Liaskos.2015b}.
These objects emerge as the combination of metasurfaces\textendash a
discipline of Physics-and nanonetworks, offering novel potential in
the interaction with electromagnetic waves.

Metasurfaces are planar, artificially structured objects that can
interact with impinging electromagnetic waves even in ways not found
in nature~\cite{Holloway.2012}. They commonly comprise a two-dimensional
pattern of a conductive material, the\emph{ meta-atom}, repeated periodically
over a dielectric substrate. The form and dimensions of the meta-atom
exemplary define the reflection angle-even at negative values-and
absorption degree of impinging waves. The \emph{HyperSurface} concept
takes the metasurface concept one step further: it is a hardware platform
that allows for forming custom meta-atoms over it, with local or global
periodicity~\cite{TheVISORSURFproject.2017}. A nanonetwork embedded
within the HyperSurface acts as the meta-atom control factor. It receives
external commands and ``draws'' meta-atoms by altering the local
conductivity of the HyperSurface accordingly.

The present paper treats the HyperSurface as a network-controlled
system, and proposes a novel protocol (Hyper-CP) for its internal
operation. The protocol can carry sensed information to the external
world for processing, as well as HyperSurface configuration commands
from the external world to the HyperSurface. Hyper-CP relies on Lyapunov
drift analysis to provide natural path-diversity to avoid congested
or out-of-power nanonetwork areas~\textcolor{black}{\cite{Georgiadis.2005}}.
Simulations are employed to evaluate Hyper-CP, which consider both
the microscopic (nanonetwork operation) and the macroscopic (detecting
impinging waves) aspects of the HyperSurface. Thus, adapting the HyperSurface
electromagnetic configuration is also enabled.

The remainder of this paper is organized as follows. Section~\ref{sec:System-Model}
provides the necessary background knowledge and the system model.
Section~\ref{sec:A-Novel-Intra-Networking} details the proposed,
HyperSurface-compatible networking scheme. Evaluation via simulations
takes place in Section~\ref{sec:sim}. Related studies are given
in Section~\ref{sec:Related-Work} and the conclusion is given in
Section~\ref{sec:Conclusion}.

\section{Background and System Model}

\label{sec:System-Model}

\textbf{Background.} \emph{Metasurfaces} are the outcome of a research
direction in Physics that seeks to construct objects with engineered
electromagnetic behavior. In their most classic form, metasurfaces
are composed of a building block (meta-atom) that is periodically
repeated over a given area. When this periodic structure is treated
at a macroscopic level, the metasurface as a whole can exhibit custom
and even exotic permeability and permittivity values. This in turn
translates to custom-and even unusual-interaction between the metasurface
and impinging electromagnetic waves. For instance, a metasurface can
exhibit an unnatural, negative refractive index, causing light to
impinging waves to bend backwards, at a negative angle of refraction.
To understand the impact of this potential, we mention that a carefully
designed stack of metasurfaces, each with its own refraction angle,
was able to render an object invisible to EM waves (cloaking) by gradually
bending them around it~\cite{Iwaszczuk.2012}.
\begin{figure}[t]
\begin{centering}
\includegraphics[bb=0bp 0bp 420bp 110bp,width=1\columnwidth]{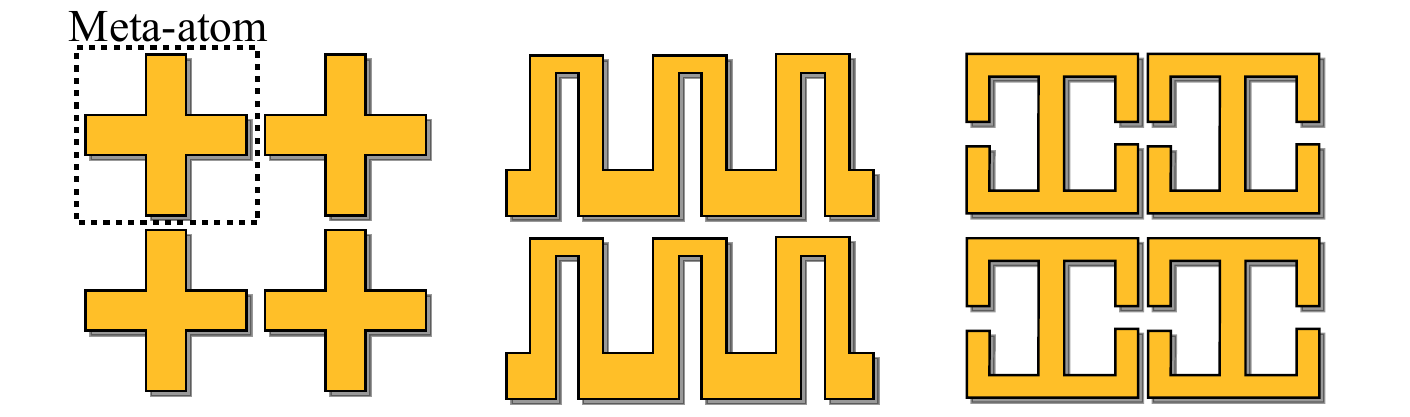}
\par\end{centering}
\caption{\label{fig:Mspatterns}Indicative meta-atom patterns that have been
employed in metasurface studies. }
\end{figure}

Some common meta-atom patterns are shown in Fig.~\ref{fig:Mspatterns}.
The meta-atom patterns are usually composed of a conductive material
and are placed over a dielectric substrate. Copper and Silicon constituted
very early material choices, while many more options, such as Graphene,
have been studied since~\cite{Zhu.2017}. One of the most commonly
studied, macroscopic interactions between a metasurface and impinging
electromagnetic waves, is wave steering~\cite{Minovich.2015}. It
expresses the reflection of an impinging wave from a given incident
angle to a direction dependent on the meta-atom configuration.
\begin{figure}[t]
\begin{centering}
\includegraphics[width=1\columnwidth]{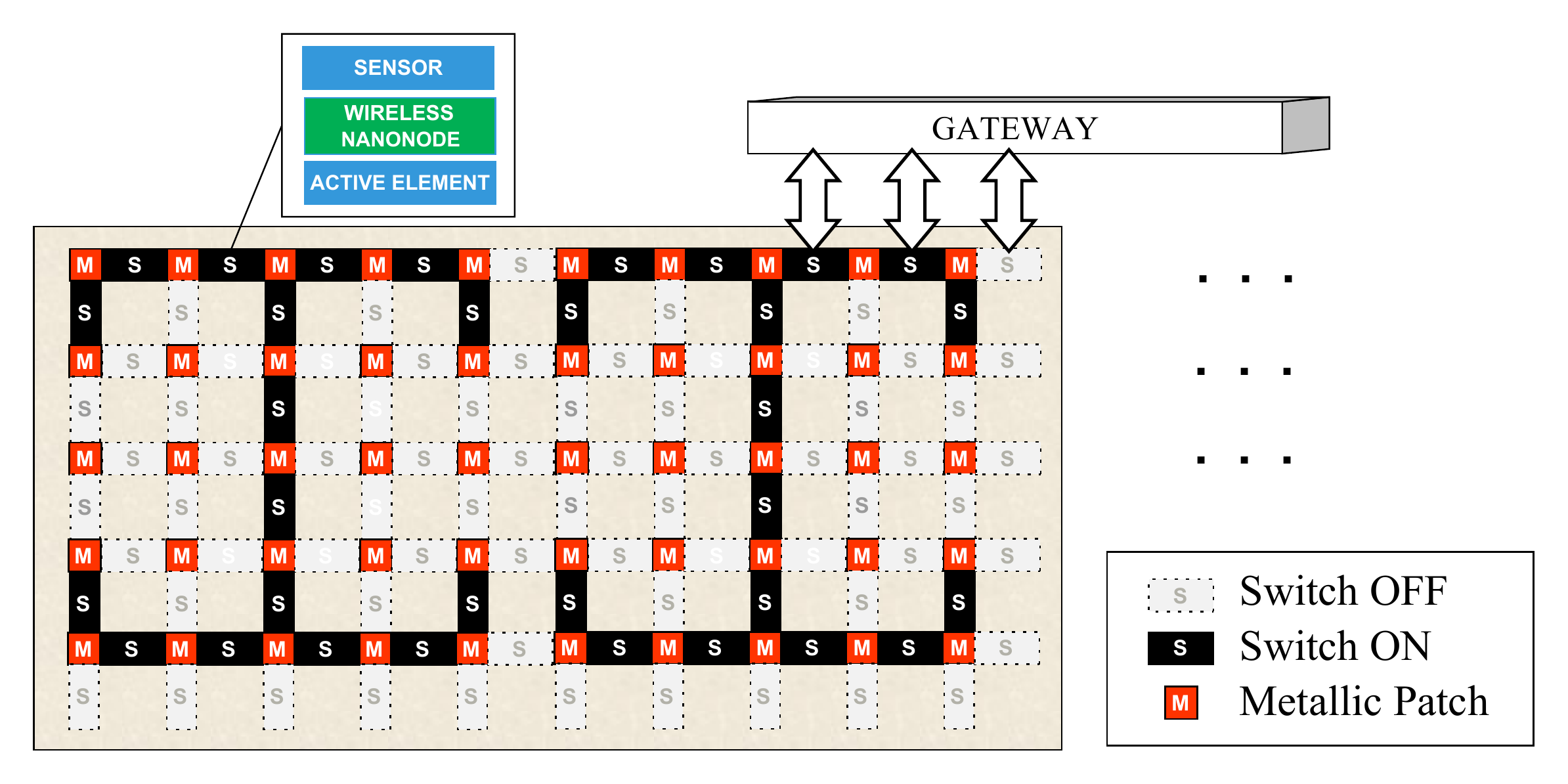}
\par\end{centering}
\caption{\label{fig:HSFControlLayer}The HyperSurface structure, comprising
passive elements (patches), active elements (binary switches) and
networked nano-controllers.}
\end{figure}

A \emph{HyperSurface} is a network-controlled metasurface. Its core
is a hardware layer that can change its internal structure to best
fit an intended meta-atom deployment. In general, this layer comprises
passive and active elements. The formation of programmatic meta-atom
arrangements is achieved by tuning the state of the active elements
accordingly. A simple example is shown in Fig.~\ref{fig:HSFControlLayer},
comprising metallic patches and CMOS switches. Custom meta-atoms are
stitched together by setting the switches appropriately to ON (conductive)
or OFF (insulating) states.

\textbf{The system model.} We assume that the state of each single
active element is managed by a nano-controller (node). Each controller
is placed under a metallic patch, to shield it from impinging eaves.
Additionally, it has wired connectivity to its immediate neighbors,
thus forming a rectangular grid network topology overall. Each node
has a hard-coded, known position over the HyperSurface in the form
of $i,\,j$ coordinates, which also act as the node's unique address.

The node power supply relies on energy-harvesting, charging a battery
of limited capacity. Each node has the ability to sense the power
level of impinging waves on the HyperSurface at their location. A
sensor measurement is captured in the form of a data packet, and is
stored on a small buffer, awaiting forwarding to the HyperSurface-external
world, via multi-hop transmission towards one or more gateways. The
gateway configuration is considered to be known by design. All nodes
are identical, and a packet is assumed to be delivered once it has
reached on of the nodes adjacent to a gateway.
\begin{figure}[t]
\begin{centering}
\includegraphics[bb=0bp 0bp 591bp 510bp,width=1\columnwidth]{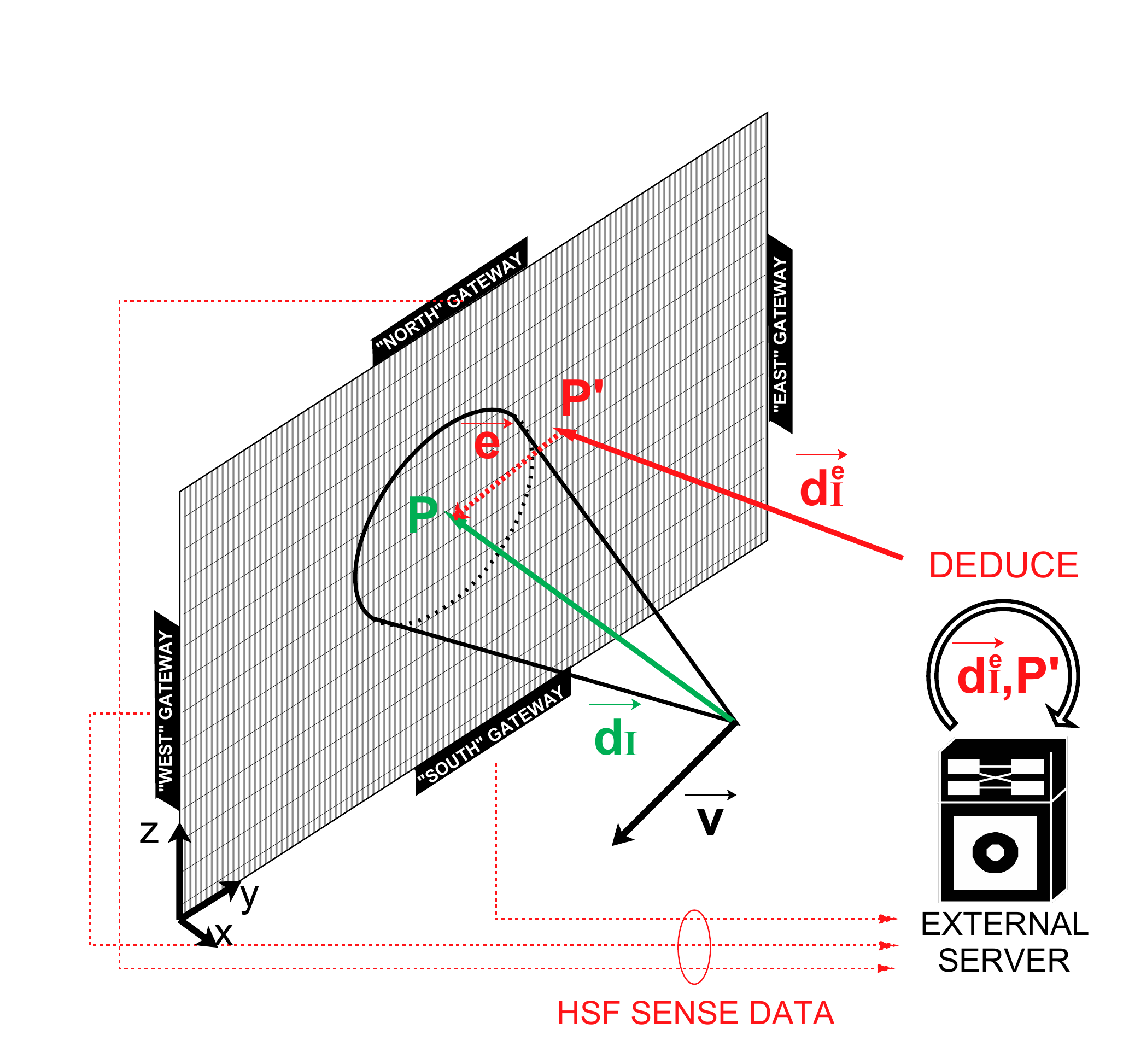}
\par\end{centering}
\caption{\label{fig:GrandPlan}HyperSurface end-functionality and interconnectivity
with external networked entities. }
\end{figure}

To facilitate the routing process, any packet about to be transmitted
(original or retransmission) carries statistics regarding the status
of the present node. Some specific statistics are discussed in Section~\ref{sec:A-Novel-Intra-Networking}.
An immediate neighbor reads this information and utilizes it to optimize
routing decisions, e.g., by extrapolating the statistics to obtain
the current neighborhood status, without exchanging specific control
packets for this task.

Figure~\ref{fig:GrandPlan} illustrates a complete HyperSurface loop
comprising a mobile wave source, and an external server that receives
the sensed data to yield the estimated, time-variant incident direction
$\overrightarrow{d_{I}^{e}}$ and impact point $P'$. Network latency
introduces a \emph{direction error}, expressed by the formation of
an angle, $\Theta^{e}$, between the real direction, $\overrightarrow{d_{I}}$,
and its estimation, $\overrightarrow{d_{I}^{e}}$ , as well as an
\emph{offset error}, $\overrightarrow{e}$, defined by the real impact
point, $P$, and the estimated, $P'$. Our goal is to study the effects
of a nanonetworking data routing protocol on the direction and offset
errors.

\section{The Hyper-CP Protocol}

\label{sec:A-Novel-Intra-Networking}

The proposed HyperSurface Control Protocol (Hyper-CP) seeks to route
a continuous stream of sensed data to the external world, taking into
account energy and latency considerations. Ideally, sensed data should
be routed towards the gateways over the shortest routes. However,
as congestion increases and node batteries get depleted, alternative
paths should be naturally followed. To accomplish these objectives,
Hyper-CP relies on the concept of Lyapunov drift minimization, which
is well-known to yield network throughput-optimality (i.e., natural
path diversity when needed), and extremely simple routing criteria~\textcolor{black}{\cite{Georgiadis.2005}}.
We complement this process by assuming outdated node status info,
and battery/latency considerations.

\textcolor{black}{The Lyapunov drift of a network is a time variant
quantity defined as~\cite{Georgiadis.2005}:
\begin{equation}
L(t)=\underset{n}{\sum}\underset{w}{\sum}U_{(n,w)}^{2}(t)
\end{equation}
where $n,\,w\ne n$ are network nodes, and $U_{(n,w)}(t)$ is the
current data buffered at node $n$ and destined towards node $w$.
We shall define an upper bound for $L(t)$ in the HyperSurface case,
and then proceed to minimize it. }

\textcolor{black}{We begin by expressing the estimated queue size
of a single node at time $t$ as:
\begin{multline}
\underset{_{(n,w)}}{U}^{e}(t)=max\left\{ 0,\underset{_{(n,w)}}{U}\left(t-\tau_{n}\right)-O_{(n,w)}^{t-\tau_{n}\to t}\right\} +\\
I_{(n,w)}^{t-\tau_{n}\to t}+G{}_{(n,w)}^{t-\tau_{n}\to t}\label{eq:strictQdynamics}
\end{multline}
where $w$ denotes a HyperSurface gateway, $\tau_{n}$ is the point
in the past when the real status of node $n$ was reported, $O_{(n,w)}^{t-\tau_{n}\to t}$,
$I_{(n,w)}^{t-\tau_{n}\to t}$ and $G_{(n,w)}^{t-\tau_{n}\to t}$
denote }\textcolor{black}{\emph{estimated}}\textcolor{black}{{} outgoing,
incoming and locally generated data (via sensing). }

\textcolor{black}{Both sides of equation~(\ref{eq:strictQdynamics})
are squared, and the following identity is applied~\cite[p. 53]{Georgiadis.2005},
\begin{equation}
V\le max\left\{ 0,\,U-\mu\right\} +A\Rightarrow V^{2}\le U^{2}+\mu^{2}+A^{2}-2U\cdot(\mu-A)\label{eq:identity}
\end{equation}
leading to:
\begin{multline}
\left(\underset{_{(n,w)}}{U}^{e}(t)\right)^{2}\le\underset{_{(n,w)}}{U}^{2}\left(t-\tau_{n}\right)+\\
\left[O_{(n,w)}^{t-\tau_{n}\to t}\right]^{2}+\left[I_{(n,w)}^{t-\tau_{n}\to t}+G_{(n,w)}^{t-\tau_{n}\to t}\right]^{2}-\\
2\cdot\underset{_{(n,w)}}{U}\left(t-\tau_{n}\right)\cdot\left[O_{(n,w)}^{t-\tau_{n}\to t}-I_{(n,w)}^{t-\tau_{n}\to t}-G_{(n,w)}^{t-\tau_{n}\to t}\right]
\end{multline}
We proceed to replace the right part of relation~(\ref{eq:strictQdynamics})
with an upper bound. Let $\mathcal{R}_{l}$ denote the nominal data
rate of a link $l$. Moreover, let $\mathcal{P}_{\left(n\to*\right)}$
and $\mathcal{P}_{\left(*\to n\right)}$ denote the set of directed
links originating from and ending at node $n$ respectively, while
also being compliant to custom selection criteria, denoted as policy
$\mathcal{P}$. Then, we write:
\begin{equation}
I_{(n,w)}^{t-\tau_{n}\to t}\le\tau_{n}\cdot\underset{l\in\mathcal{P}_{\left(*\to n\right)}}{\sum\mathcal{R}_{l}},\,\,O_{(n,w)}^{t-\tau\to t}\le\tau_{n}\cdot\underset{l\in\mathcal{P}_{\left(n\to*\right)}}{\sum\mathcal{R}_{l}}\label{eq:relax1-1}
\end{equation}
Using the bounds of (\ref{eq:relax1-1}) and setting ${\scriptstyle {\Delta U_{(n,w)}^{2}(t)=\left(\underset{_{(n,w)}}{U}^{e}(t)\right)^{2}}}$
${\scriptstyle {-\underset{_{(n,w)}}{U}^{2}\left(t-\tau_{n}\right)}}$,
we obtain:
\begin{multline}
\hspace{-5bp}\Delta U_{(n,w)}^{2}(t)\le\tau_{n}^{2}\cdot\left(\underset{l\in\mathcal{P}_{\left(n\to*\right)}}{\sum\mathcal{R}_{l}}\right)^{2}+\left[{\scriptstyle {\tau_{n}\cdot\underset{l\in\mathcal{P}_{\left(*\to n\right)}}{\sum\mathcal{R}_{l}}+G_{(n,w)}^{t-\tau_{n}\to t}}}\right]^{2}\\
-2\cdot\underset{_{(n,w)}}{U}\left(t-\tau_{n}\right)\cdot\left[\tau_{n}\cdot\underset{l\in\mathcal{P}_{\left(n\to*\right)}}{\sum\mathcal{R}_{l}}-\tau_{n}\cdot\underset{l\in\mathcal{P}_{\left(*\to n\right)}}{\sum\mathcal{R}_{l}}-G_{(n,w)}^{t-\tau_{n}\to t}\right]\label{eq:RHS_raw}
\end{multline}
Inequality (\ref{eq:RHS_raw}) can be rewritten as follows:
\begin{multline}
\Delta U_{(n,w)}^{2}(t)\le\left[\tau_{n}\cdot\underset{l\in\mathcal{P}_{\left(*\to n\right)}}{\sum\mathcal{R}_{l}}+\underset{_{(n,w)}}{U}\left(t-\tau_{n}\right)+G{}_{(n,w)}^{t-\tau_{n}\to t}\right]^{2}\\
+\left[\tau_{n}\cdot\underset{l\in\mathcal{P}_{\left(n\to*\right)}}{\sum\mathcal{R}_{l}}-\underset{_{(n,w)}}{U}\left(t-\tau_{n}\right)\right]^{2}-2\cdot\underset{_{(n,w)}}{U}^{2}\left(t-\tau_{n}\right)
\end{multline}
We sum both sides $\forall n,w$, to obtain the Lyapunov drift at
the left part:
\begin{multline}
\Delta L(t)\le\underset{\forall n}{\sum}\underset{\forall w}{\sum}\underset{}{\left[{\scriptstyle {\tau_{n}\cdot\underset{l\in\mathcal{P}_{\left(*\to n\right)}}{\sum\mathcal{R}_{l}}+\underset{_{(n,w)}}{U}\left(t-\tau_{n}\right)+G_{(n,w)}^{t-\tau_{n}\to t}}}\right]^{2}}\\
+\underset{\forall n}{\sum}\underset{\forall w}{\sum}\underset{}{\left[{\scriptstyle {\tau_{n}\cdot\underset{l\in\mathcal{P}_{\left(n\to*\right)}}{\sum\mathcal{R}_{l}}-\underset{_{(n,w)}}{U}\left(t-\tau_{n}\right)}}\right]^{2}}-2\cdot\underset{\forall n}{\sum}\underset{\forall w}{\sum}\underset{_{(n,w)}}{U}^{2}\left(t-\tau_{n}\right)\label{eq:earlyInsights}
\end{multline}
}The terms $\sum\mathcal{R}_{l}$ in the right part can be viewed
as the data routing decisions to be taken at time $t$. The following
Theorem provides the optimal decisions that minimize the Lyapunov
drift bound, i.e., the right part of relation (\ref{eq:earlyInsights}).

\textbf{Theorem.} T\textcolor{black}{he Lyapunov bound-optimizing
routing decision at node $n$, regarding traffic destined to gateway
$w$, is to forward the traffic to neighboring node $m^{*}$:
\begin{equation}
m^{*}=argmax_{m\in\mathcal{M}}\left\{ \underset{_{(n,w)}}{U}(t)-\left(\underset{_{(m,w)}}{U}^{e}(t)+G_{(m,w)}^{t-\tau_{m}\to t}\right)\right\} \label{eq:foresightOptimal}
\end{equation}
where $\mathcal{M}$ is the set containing the neighbors of $n$ that
comply with policy $\mathcal{P}$. }

The proof is omitted.

\begin{algorithm}[t]
\textbf{\textcolor{black}{\footnotesize{}1: For each}}\textcolor{black}{\footnotesize{}
gateway $w$}{\footnotesize \par}

\textbf{\textcolor{black}{\footnotesize{}2: }}\textcolor{black}{\footnotesize{}$m^{*}\gets\underset{m\in\mathcal{M}}{argmax}\left\{ \underset{_{(n,w)}}{U}(t)-\left(\underset{_{(m,w)}}{U}^{e}(t)+G_{(m,w)}^{t-\tau_{m}\to t}\right)\right\} $}{\footnotesize \par}

\textbf{\textcolor{black}{\footnotesize{}3:}}\textcolor{black}{\footnotesize{}
Forward all packets for $w$ to $m^{*}$ }{\footnotesize \par}

\textbf{\textcolor{black}{\footnotesize{}4: end for}}{\footnotesize \par}

\textbf{\textcolor{black}{\footnotesize{}5: goto}}\textcolor{black}{\footnotesize{}
1 }{\footnotesize \par}

\textcolor{black}{\footnotesize{}\caption{\label{alg:FBPR}The Hyper-CP packet forwarding loop.}
}{\footnotesize{} }{\footnotesize \par}
\end{algorithm}

\textcolor{black}{Theorem~$1$ dictates a very simple data forwarding
loop for the nano-nodes, formalized as Algorithm~\ref{alg:FBPR},
which is repeated as long as a node is powered and has pending packets
in its queue. A node $n$ needs just an estimation of its neighbors
$\underset{_{(m,w)}}{U}^{e}(t)$ and $G_{(m,w)}^{t-\tau_{m}\to t}$
quantities. The latter is the packet generation rate at a node, due
to its sensing duties. Assuming a global, hard-wired sense rate of
$\overline{g}$ (packets per time unit), a node $n$ can estimate
$G_{(m,w)}^{t-\tau_{m}\to t}$ for a neighbor $m$ as $\overline{g}\cdot\tau_{m}$.
To estimate $\underset{_{(m,w)}}{U}^{e}(t)$, however, node $n$ requires
some statistics, mentioned in the system model (Section~\ref{sec:System-Model}).
Particularly, the packet at $t-\tau_{m}$ that reached node $n$ from
$m$, reports not only $\underset{_{(m,w)}}{U}(t-\tau_{m})$, but
also its gradient,~$\overline{u}$. Thus, $\underset{_{(m,w)}}{U}^{e}(t)$
is estimated as $max\left\{ 0,\,\underset{_{(m,w)}}{U}(t-\tau_{m})+\overline{u}\cdot\tau_{m}\right\} $. }

\textcolor{black}{Similarly, node $n$ can obtain an estimate of the
battery level of node $m$, $B_{m}^{e}(t)$. This information, coupled
with the node's coordinates $\left\{ i_{m},j_{m}\right\} $,~$\left\{ i_{n},j_{n}\right\} $
can be employed in a battery and latency-aware policy $\mathcal{P}$.
Specifically, the policy specifies the valid neighbors $\mathcal{M}$
of $n$, as the ones that:}
\begin{itemize}
\item Have a battery level greater than a customized threshold: \textcolor{black}{$B_{m}^{e}(t)\ge B_{TH}$.}
\item Are nearer to the intended gateway $w$ than node $n$:
\begin{equation}
\left|i_{w}-i_{m}\right|+\left|j_{w}-j_{m}\right|<\left|i_{w}-i_{n}\right|+\left|j_{w}-j_{n}\right|\label{eq:late}
\end{equation}
\end{itemize}
Condition (\ref{eq:late}) can be ignored when the battery constraint
is not met. Finally, by convention, in absence of node statistics
batteries are assumed to be fully charged and packet buffers empty.
When a node loses power, its estimations are also reset to these default
values.

\section{Evaluation }

\label{sec:sim}

A HyperSurface is simulated in the Anylogic platform \cite{XJTechnologies.2013},
considering both the microscopic (nanonetwork operation) and the macroscopic
(interaction with impinging waves) aspects:

\textbf{Macroscopic setup}. We assume the system model of Fig.~\ref{fig:GrandPlan}.
The HyperSurface has dimensions $Y_{max}=500$~cm, $Z_{max}=500$~cm.
A wave point-source enters at point $\left\{ 500\right.$, $0$, $\left.500\right\} $~cm
and moves towards point $\left\{ 500,500,0\right\} $, parallel to
the $y-z$ plane with constant velocity $\left|\overrightarrow{v}\right|=5$~cm/sec.
Its emission cone has a direction perpendicular to the $y-z$ plane,
and an opening angle of $10^{o}$. The emission is isotropic within
this cone, and has a constant power of $100$~mW. Free-space path
loss is the sole attenuation factor.

\textbf{Microscopic setup}. We consider a $31\times31$ rectangular
grid of asynchronous nano-nodes. The four nodes placed at the middle-most
HyperSurface edge points act as gateways and have no power constraints.
All other nodes have a power harvesting rate of $10$ pJ/sec, an energy
drain per (\emph{wired}) packet emission of $10$~pJ, and a battery
capacity of $1000$~pJ~\cite{Jornet.2012b}. Each node senses the
impinging power periodically, (input variable\emph{ sensing period}).
Zero-power measurements are ignored and do not propagate within the
network. Original packets are sent towards their nearest gateway.
The total packet propagation time over any link is set to $1$~msec.
The battery threshold $B_{TH}$ is set to $50\%$ in all cases, but
is disregarded if no neighboring node can meet it. Each node has a
packet queue that can hold up to $10$ packets before overflowing.
As a means against power depletion, once a packet queue empties completely,
the corresponding node enters a ``resting'' mode for $1$ sec, refraining
from further packet transmissions to replenish his battery level.

A packet that has reached a gateway is considered to be available
to an external server. The server constantly derives the estimated
incident point $P'$ as the average of the coordinates of the last
$50$ sensors with a power reading. Moreover, the estimated incident
direction $\overrightarrow{d_{I}^{e}}$ is continuously calculated
as the conic section that best fits the last $50$ power measurements.

Two Hyper-CP variants are compared. The first is given perfect knowledge
of the nodes' statuses (\emph{ideal}), while the second relies on
the \emph{estimations} described in Section~\ref{sec:A-Novel-Intra-Networking}.
The macroscopic performance metrics of interest are the average offset
and direction errors throughout the source movement. Microscopic metrics
are the corresponding average node battery and packet overflow rate.
All runs are repeated $100$ times and mean values are presented in
the ensuing Figures.

\begin{figure}[t]
\begin{centering}
\includegraphics[width=1\columnwidth]{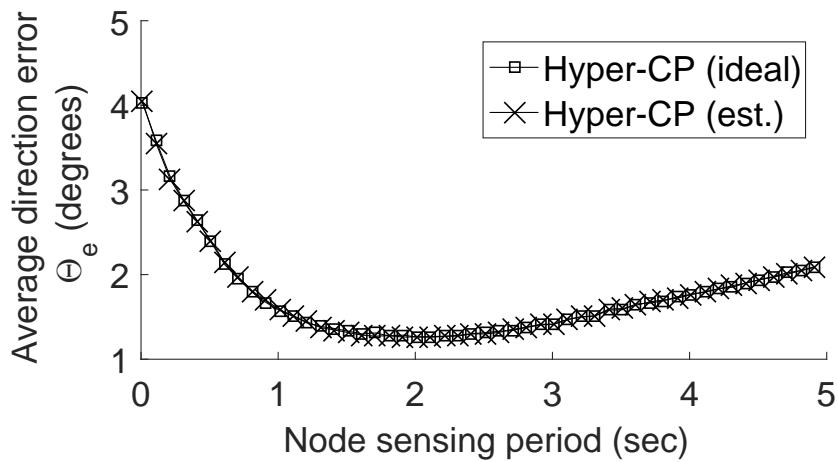}
\par\end{centering}
\caption{\label{fig:direction}Hyper-CP direction error versus the node sensing
period.}
\end{figure}

\begin{figure}[t]
\begin{centering}
\includegraphics[width=1\columnwidth]{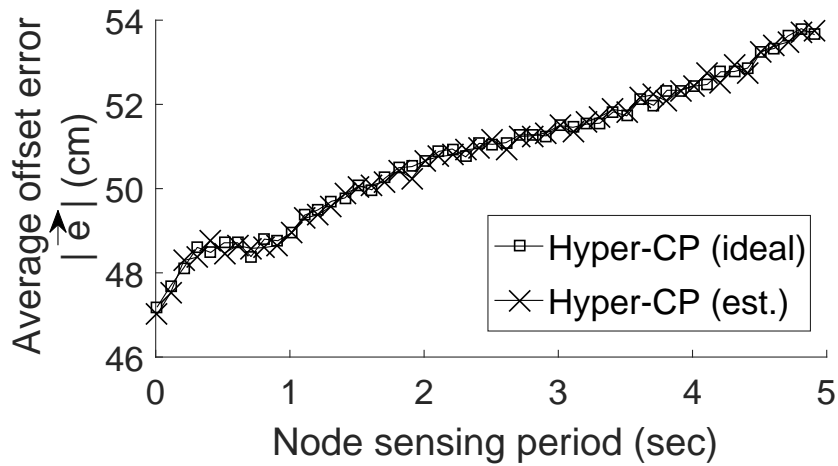}
\par\end{centering}
\caption{\label{fig:offset}Hyper-CP offset error versus the node sensing period.}
\end{figure}

\begin{figure}[t]
\begin{centering}
\includegraphics[width=1\columnwidth]{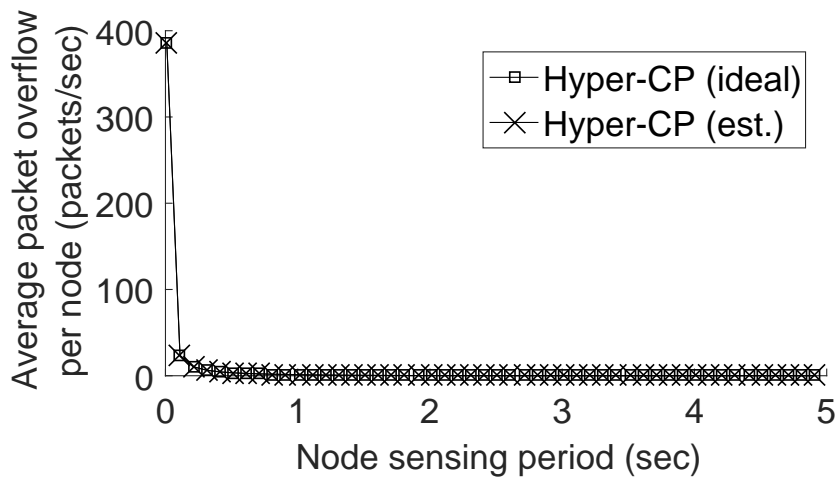}
\par\end{centering}
\caption{\label{fig:overflow}Node packet overflow rate versus the sensing
period.}
\end{figure}

\begin{figure}[t]
\begin{centering}
\includegraphics[width=1\columnwidth]{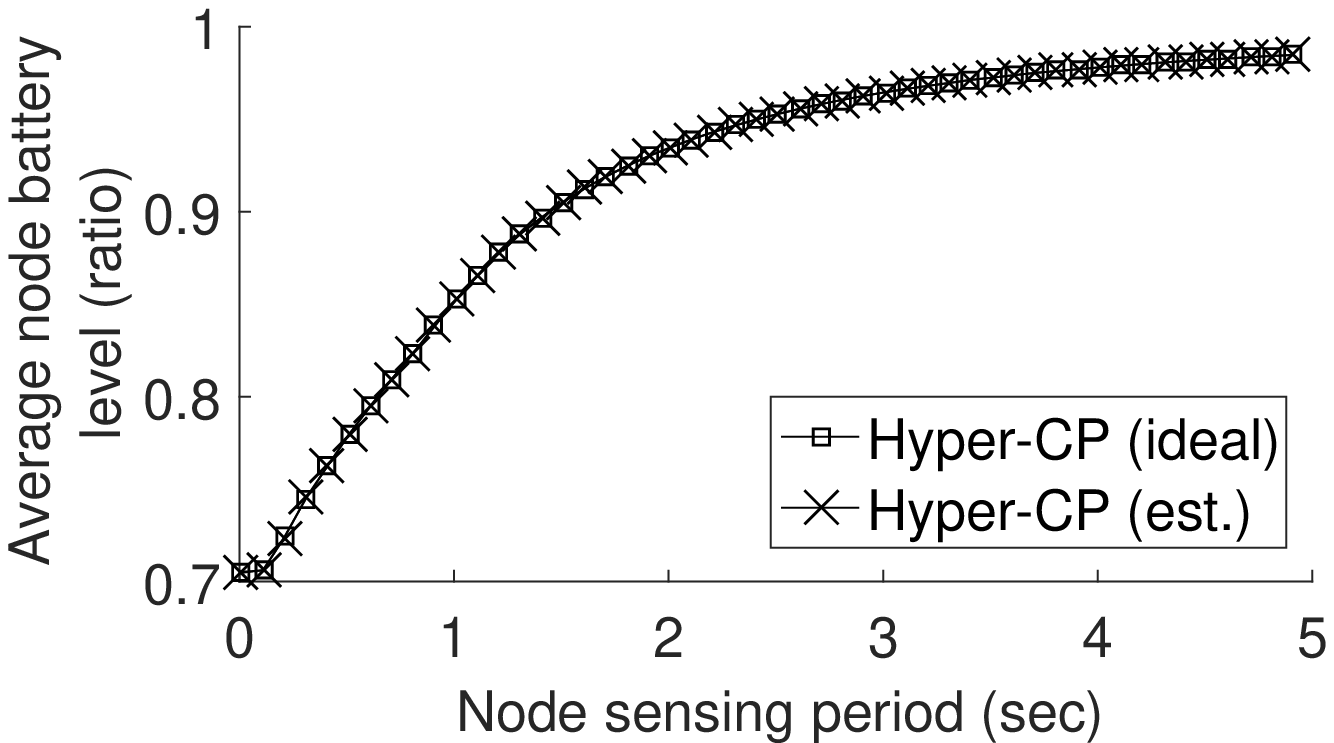}
\par\end{centering}
\caption{\label{fig:battery}Average node battery versus the node sensing period.}
\end{figure}

\textbf{Results}. The macroscopic-aspect results are presented in
Fig.~\ref{fig:direction} and \ref{fig:offset}, while the corresponding
microscopic-aspect performance is shown in Fig.~\ref{fig:overflow}
and \ref{fig:battery}.

The x-axis in all Figures represents the node sense period, with separate
experiments ran for $0.01$ to $5$ seconds between successive node
sensing events. Intuitively, the sampling period should affect both
the macroscopic and microscopic performance. On one hand, a smaller
sensing period means more frequent power sampling, thus facilitating
the tracking of a moving source. On the other hand, it also leads
to an increasing number of generated packets, which need to be routed
towards the gateways. This might in turn cause congestion, negatively
affecting the macroscopic performance.

The described trade-off is evident in the case of the direction error
in Fig.~\ref{fig:direction}. For sampling periods of $0.01$ to
$\sim1$ sec, the network is in a congested state, as shown by the
packet overflow rate in Fig.~\ref{fig:overflow}. The high number
of generated packets leads to full packet buffers at the nodes and,
thus, to increased buffering times and packet drops, negatively affecting
the direction error in Fig.~\ref{fig:direction}. As the sampling
period increases, the network stabilizes (Fig.~\ref{fig:overflow},
periods $1\to3$ sec) and the direction error reaches a minimal value
of approximately $1^{o}$. Further increase in the sampling period
exhibits a linear increase in the direction error, since the node
measurements are not frequent enough to keep track of the moving source.
We note that the direction error is an angle metric, and thus generally
sensitive to $P'$ point estimations. Nonetheless, the estimated direction
is generally parallel to the real one, exhibiting a maximum error
of $4^{o}$.

The offset error, shown in Fig.~\ref{fig:offset}, exhibits a linear
increase versus the sampling period, and is not affected by a severe
network congestion. The offset error is expressed as the norm of the
displacement vector, $\overrightarrow{e}$, and thus does not contain
error angle information. Moreover, the metric is derived by fitting
the measurements incoming to the gateways to the best conic section,
as described above. A very limited amount of node measurements is
required for this task, and, therefore, a highly congested network
state does not have a strong impact on the metric. Thus, the offset
error exhibits only a linear increase with the sampling period. Notice
that the offset error is moderately high, i.e., $\sim50$ cm, which
is the $10\%$ of the overall space dimensions. However, notice that
in conjunction with the minimal direction error discussed above, the
estimated source position can be considered as exhibiting a constant
displacement with regard to the real position. This means that the
external server can employ mobility prediction schemes to mitigate
the total macroscopic error.

The average node battery levels, shown in Fig.~\ref{fig:battery},
are naturally expected; smaller sensing periods lead to a higher number
of packets to be routed, and eventually to a greater power expense
(lower average battery). The node resting mode effectively guards
against network energy depletion: The minimum average battery level
is $70\%$ for the minimum sampling period of $0.01$ sec. Moreover,
it increases asymptotically towards $100\%$ for higher sensing periods
which naturally stress the network less.

Finally, the Hyper-CP protocol is shown to operate equally well with
ideal and estimated knowledge of the nodes' status. Queue size estimations
may have a negligible effect in the routing decision of relation (\ref{eq:foresightOptimal}).
Assuming that the estimation error is constant (on average) for all
nodes, the decision of relation (\ref{eq:foresightOptimal}) is not
significantly affected by it. Additionally, we note that the grid
topology also works in favor of the estimations. Each node has a limited
number of neighbors (i.e., $4$) in the studied grid layout. Moreover,
condition (\ref{eq:late}) may limit the considered nodes further.
Thus, routing decision may refer to a limited number of neighbors
(e.g., $2$) reducing the impact of inaccurate estimations. Wireless
network setups, where each node can have multiple neighbors, can constitute
a future work direction. The wireless approach can impose a greater
energy power consumption to the HyperSurface controller network, but
provides the potential for versatility to the HyperSurface deployment,
compared to the studied wired approach.

\section{Related Work}

\label{sec:Related-Work}

In this Section we review works studying the limitations of nanonetworking
in general and related nano-communication protocols.

Wang et al. study manufacturing limitations at nano-scale and highlight
the need for unique nano-addressing and nano-routing solutions~\cite{Wang.2013}.
First, the power supply units of autonomous nano-nodes can scavenge
energy for $1$ packet transmission per approximately $10$ sec~\cite{Jornet.2012b}.
Second, a class of wireless nano-communication modules are expected
to operate at the $T$Hz band, which translates to highly lossy channel
conditions due to acute molecular absorption phenomena~\cite{Jornet.2011}.
Third, manufacturing restrictions and cost considerations correspond
to ``weak'' nano-node hardware, i.e., limited CPU and data storage
capabilities~\cite{Akyildiz.2010b,Tairin.2017}. The impact of these
restrictions on nano-node addressing is that assigning unique identifiers
to each node is not scalable, mainly due to power restrictions. Regarding
data routing, on the other hand, the expectedly frequent transmission
failures require a mechanism to balance path redundancy and energy
consumption, while incurring low computational complexity and memory
overhead~\cite{Wang.2013}. Subsequently, several Medium Access Control
(MAC) solutions have been proposed, which assume mobile nodes and
$T$Hz-operating radio modules~\cite{Mohrehkesh.2015,Yao.2016}.
These specifications\textendash apart from the strong emphasis on
energy efficiency\textendash are not aligned to the described HyperSurface
nanonetworks.

Regarding the authors' prior work, a ray-tracing-based simulation
technique for nanonetworks was initially studied and validated against
precise computational techniques (Finite-Differences Time-Domain)
in~\cite{Kantelis.2014}. Subsequent work focused on application
setups were packet flooding constitutes a good match. Liaskos et al.
proposed a nature-inspired self-organizing nanonetwork architecture~\cite{Liaskos.2015}.
Tsioliaridou et al. worked towards minimal complexity schemes, expressed
by the assumptions of nano-CPUs supporting only basic integer arithmetic,
and few bytes of memory~\cite{Tsioliaridou.2016}. A joint peer-to-peer
routing and geo-addressing approach for 2D networks was proposed in~\cite{Tsioliaridou.2015}.
The routing component operated via selective flooding contained within
network sub-areas (reducing the number of involved retransmitters
per communicating node-pair), and with and extremely low computational
footprint. The work was then extended to operate in 3D network layouts~\cite{Tsioliaridou.2016c},
and by proposing a routing approach over curvilinear paths, essentially
minimizing the required retransmitters~\cite{Tsioliaridou.2017}.
The present work differentiates by proposing a control protocol specifically
for the HyperSurface scenario. The metric of performance is the efficient
detection of the direction-of-arrival for waves impinging on the HyperSurface.
Moreover, the protocol is point-to-point based, while it considers
energy harvesting power supply at the nano-nodes.

\section{Conclusion}

\label{sec:Conclusion}

The present work studied nanonetwork-controlled metasurfaces, the
HyperSurfaces, which can sense their environment and change their
electromagnetic behavior according to external, programmatic commands.
A protocol that relays such information must offer natural path diversity,
depending on congestion, and nano-node energy shortage. The study
proposed the Hyper-CP protocol, which was shown to uphold these conditions
via simulations. The Hyper-CP protocol was derived analytically, using
a Lyapunov drift minimization approach.

 \bibliographystyle{acm}

\begin{thebibliography}{10}

\bibitem{Akyildiz.2010c}
{\sc Akyildiz, I., and Jornet, J.}
\newblock {The Internet of nano-things}.
\newblock {\em {IEEE Wireless Communications} 17}, 6 (2010), 58--63.

\bibitem{Akyildiz.2010b}
{\sc Akyildiz, I.~F., and Jornet, J.~M.}
\newblock {Electromagnetic wireless nanosensor networks}.
\newblock {\em {Nano Communication Networks} 1}, 1 (2010), 3--19.

\bibitem{Georgiadis.2005}
{\sc Georgiadis, L., Neely, M.~J., and Tassiulas, L.}
\newblock {Resource Allocation and Cross-Layer Control in Wireless Networks}.
\newblock {\em {FNT in Networking (Foundations and Trends in Networking)} 1}, 1
  (2005), 1--144.

\bibitem{Holloway.2012}
{\sc Holloway, C.~L., Kuester, E.~F., Gordon, J.~A., O'Hara, J., Booth, J., and
  Smith, D.~R.}
\newblock {An Overview of the Theory and Applications of Metasurfaces: The
  Two-Dimensional Equivalents of Metamaterials}.
\newblock {\em {IEEE Antennas and Propagation Magazine} 54}, 2 (2012), 10--35.

\bibitem{Iwaszczuk.2012}
{\sc Iwaszczuk, K., Strikwerda, A.~C., Fan, K., Zhang, X., Averitt, R.~D., and
  Jepsen, P.~U.}
\newblock {Flexible metamaterial absorbers for stealth applications at
  terahertz frequencies}.
\newblock {\em {Optics Express} 20}, 1 (2012), 635.

\bibitem{Jornet.2011}
{\sc Jornet, J., and Akyildiz, I.}
\newblock {Channel Modeling and Capacity Analysis for Electromagnetic Wireless
  Nanonetworks in the Terahertz Band}.
\newblock {\em {IEEE Transactions on Wireless Communications} 10}, 10 (2011),
  3211--3221.

\bibitem{Jornet.2012b}
{\sc Jornet, J., and Akyildiz, I.}
\newblock {Joint Energy Harvesting and Communication Analysis for Perpetual
  Wireless Nanosensor Networks in the Terahertz Band}.
\newblock {\em {IEEE Transactions on Nanotechnology} 11}, 3 (2012), 570--580.

\bibitem{Kantelis.2014}
{\sc Kantelis, K., Amanatiadis, S., Liaskos, C., Kantartzis, N., Konofaos, N.,
  Nicopolitidis, P., and Papadimitriou, G.}
\newblock {On the Use of FDTD and Ray-Tracing Schemes in the Nanonetwork
  Environment}.
\newblock {\em {IEEE Communications Letters} 18}, 10 (2014), 1823--1826.

\bibitem{Liaskos.2015}
{\sc Liaskos, C., and Tsioliaridou, A.}
\newblock {A Promise of Realizable, Ultra-Scalable Communications at
  nano-Scale: A multi-Modal nano-Machine Architecture}.
\newblock {\em {IEEE Transactions on Computers} 64}, 5 (2015), 1282--1295.

\bibitem{Liaskos.2015b}
{\sc Liaskos, C., Tsioliaridou, A., Pitsillides, A., Akyildiz, I.~F.,
  Kantartzis, N., Lalas, A., Dimitropoulos, X., Ioannidis, S., Kafesaki, M.,
  and Soukoulis, C.}
\newblock {Design and Development of Software Defined Metamaterials for
  Nanonetworks}.
\newblock {\em {IEEE Circuits and Systems Mag.} 15}, 4 (2015), 12--25.

\bibitem{Minovich.2015}
{\sc Minovich, A.~E., Miroshnichenko, A.~E., Bykov, A.~Y., Murzina, T.~V.,
  Neshev, D.~N., and Kivshar, Y.~S.}
\newblock {Functional and nonlinear optical metasurfaces: Optical
  metasurfaces}.
\newblock {\em {Laser {\&} Photonics Reviews} 9}, 2 (2015), 195--213.

\bibitem{Mohrehkesh.2015}
{\sc Mohrehkesh, S., Weigle, M.~C., and Das, S.~K.}
\newblock {DRIH-MAC: A Distributed Receiver-Initiated Harvesting-Aware MAC for
  Nanonetworks}.
\newblock {\em {Molecular, Biological and Multi-Scale Communications, IEEE
  Transactions on} 1}, 1 (2015), 97--110.

\bibitem{Piro.2015}
{\sc Piro, G., Boggia, G., and Grieco, L.}
\newblock {On the design of an energy-harvesting protocol stack for Body Area
  Nano-NETworks}.
\newblock {\em {Nano Communication Networks} 6}, 2 (2015), 74--84.

\bibitem{Tairin.2017}
{\sc Tairin, S., Nurain, N., and {Islam,A.~B.~M.~Alim~Al}}.
\newblock {Network-level performance enhancement in wireless nanosensor
  networks through multi-layer modifications}.
\newblock In {\em {2017 International Conference on Networking, Systems and
  Security (NSysS)}}, pp.~75--83.

\bibitem{TheVISORSURFproject.2017}
{\sc {The~VISORSURF~project}}.
\newblock {A Hardware Platform for Software-driven Functional Metasurfaces}.
\newblock {\em {Horizon 2020 Future Emerging Technologies}\/} (2017).

\bibitem{Tsioliaridou.2017}
{\sc Tsioliaridou, A., Liaskos, C., Dedu, E., and Ioannidis, S.}
\newblock {Packet Routing in 3D Nanonetworks: A Lightweight, Linear-path
  Scheme}.
\newblock {\em {Nano Communication Networks} PP\/} (2017), PP.

\bibitem{Tsioliaridou.2015}
{\sc Tsioliaridou, A., Liaskos, C., Ioannidis, S., and Pitsillides, A.}
\newblock {CORONA: A Coordinate and Routing system for Nanonetworks}.
\newblock In {\em {ACM NANOCOM'15}\/} (2015), p.~18.

\bibitem{Tsioliaridou.2016}
{\sc Tsioliaridou, A., Liaskos, C., Ioannidis, S., and Pitsillides, A.}
\newblock {Lightweight, self-tuning data dissemination for dense nanonetworks}.
\newblock {\em {Nano Communication Networks (Special Issue on EM Nanonetworks)}
  8\/} (2016), 2--15.

\bibitem{Tsioliaridou.2016c}
{\sc Tsioliaridou, A., Liaskos, C., Pachis, L., Ioannidis, S., and Pitsillides,
  A.}
\newblock {N3: Addressing and Routing in 3D Nanonetworks}.
\newblock In {\em {IEEE ICT'16}\/} (2016), pp.~1--6.

\bibitem{Wang.2013}
{\sc Wang, P., Jornet, J.~M., {Abbas~Malik}, M., Fadel, E., and Akyildiz,
  I.~F.}
\newblock {Energy and spectrum-aware MAC protocol for perpetual wireless
  nanosensor networks in the Terahertz Band}.
\newblock {\em {Ad Hoc Networks} 11}, 8 (2013), 2541--2555.

\bibitem{XJTechnologies.2013}
{\sc {XJ~Technologies}}.
\newblock {\em {The AnyLogic Simulator}}.
\newblock 2013.

\bibitem{Yao.2016}
{\sc Yao, X.-W., and Jornet, J.~M.}
\newblock {TAB-MAC: Assisted beamforming MAC protocol for Terahertz
  communication networks}.
\newblock {\em {Nano Communication Networks} 9\/} (2016), 36--42.

\bibitem{Zhu.2017}
{\sc Zhu, A.~Y., Kuznetsov, A.~I., Luk'yanchuk, B., Engheta, N., and Genevet,
  P.}
\newblock {Traditional and emerging materials for optical metasurfaces}.
\newblock {\em {Nanophotonics} 6}, 2 (2017).

\end{thebibliography}

\end{document}